\input harvmac
\input amssym

\def\IR{\Bbb{R}}

 \def\cM{{\cal M}}
\def\cN{{\cal N}} 
\def\cP{{\cal P}} 
\def\cR{{\cal R}} 
 
 \def\cW{{\cal W}}
 
\def\oneone{\rlap 1\mkern4mu{\rm l}}
%

\def\sst#1{{\scriptscriptstyle #1}}
\def\ft#1#2{{\textstyle{{\scriptstyle #1}\over {\scriptstyle #2}}}}
\def\fft#1#2{{#1 \over #2}}
\def\ep{\epsilon}

\def\td{\tilde}
\def\3{{\sst{(3)}}}
\def\4{{\sst{(4)}}}
\def\tX{{{\wtd X}}} 
\def\wtd{\widetilde}

\def\coeff#1#2{\relax{\textstyle {#1 \over #2}}\displaystyle}
\def\half{{1 \over 2}}

\def\nup#1({Nucl.\ Phys.\ $\us {B#1}$\ (}
\def\plt#1({Phys.\ Lett.\ $\us  {#1B}$\ (}
\def\plt#1({Phys.\ Lett.\ $\us  {#1B}$\ (}
\def\cmp#1({Comm.\ Math.\ Phys.\ $\us  {#1}$\ (}
\def\prp#1({Phys.\ Rep.\ $\us  {#1}$\ (}
\def\prl#1({Phys.\ Rev.\ Lett.\ $\us  {#1}$\ (}
\def\prv#1({Phys.\ Rev.\ $\us  {#1}$\ (}
\def\mpl#1({Mod.\ Phys.\ Let.\ $\us  {A#1}$\ (}
\def\ijmp#1({Int.\ J.\ Mod.\ Phys.\ $\us{A#1}$\ (}
\def\atmp#1({Adv.\ Theor.\ Math.\ Phys.\ $\bf {#1}$\ (}
\def\cqg#1({Class.\ Quant.\ Grav.\ $\bf {#1}$\ (}
\def\jag#1({Jour.\ Alg.\ Geom.\ $\us {#1}$\ (}
\def\jhep#1({JHEP $\bf {#1}$\ (}

%

%
%
\lref\GauntlettSC{J.~P.~Gauntlett, D.~Martelli, S.~Pakis and 
D.~Waldram,``G-structures and wrapped NS5-branes,'' hep-th/0205050.
%
}
%
\lref\GauntlettRV{J.~P.~Gauntlett, N.~Kim, S.~Pakis and D.~Waldram,
``M-theory solutions with AdS factors,''
Class.\ Quant.\ Grav.\  {\bf 19}, 3927 (2002), hep-th/0202184.
}
%
\lref\GauntlettNW{J.~P.~Gauntlett, J.~B.~Gutowski, C.~M.~Hull, 
S.~Pakis and H.~S.~Reall,
``All supersymmetric solutions of minimal supergravity in five dimensions,''
hep-th/0209114.
}
%
\lref\GauntlettFZ{J.~P.~Gauntlett and S.~Pakis,
``The geometry of D = 11 Killing spinors,''
hep-th/0212008.
}
%
\lref\deWitIG{
B.~de Wit and H.~Nicolai,
``N=8 Supergravity,''
Nucl.\ Phys.\ B {\bf 208}, 323 (1982).
}
\lref\CremmerUP{   
E.~Cremmer and B.~Julia,
``The SO(8) Supergravity,''
Nucl.\ Phys.\ B {\bf 159}, 141 (1979).
}
\lref\BakXX{D.~s.~Bak and A.~Karch, ``Supersymmetric 
brane-antibrane configurations,''Nucl.\ Phys.\ B 
{\bf 626}, 165 (2002), hep-th/0110039.0.
%
}
%
\lref\CorradoWX{
R.~Corrado, M.~Gunaydin, N.~P.~Warner and M.~Zagermann,
``Orbifolds and flows from gauged supergravity,''
Phys.\ Rev.\ D {\bf 65}, 125024 (2002), hep-th/0203057.
%
}
%
\lref\MateosQS{
D.~Mateos and P.~K.~Townsend,
``Supertubes,''
Phys.\ Rev.\ Lett.\  {\bf 87}, 011602 (2001), hep-th/0103030.
}
%
\lref\PilchUE{
K.~Pilch and N.~P.~Warner,
``N = 2 supersymmetric RG flows and the IIB dilaton,''
Nucl.\ Phys.\ B {\bf 594}, 209 (2001), hep-th/0004063.
}
%
\lref\CveticAU{
M.~Cvetic, H.~Lu and C.~N.~Pope,
``Four-dimensional N = 4, SO(4) gauged supergravity from D = 11,''
Nucl.\ Phys.\ B {\bf 574}, 761 (2000), hep-th/9910252.
}
%
%
\lref\deWitNZ{
B.~de Wit, H.~Nicolai and N.~P.~Warner,
Nucl.\ Phys.\ B {\bf 255}, 29 (1985).
}
%
\lref\deWitIY{
B.~de Wit and H.~Nicolai,
Nucl.\ Phys.\ B {\bf 281}, 211 (1987).
}
%
\lref\DonagiCF{
R.~Donagi and E.~Witten,
Nucl.\ Phys.\ B {\bf 460}, 299 (1996)
[arXiv:hep-th/9510101].
}
%

\Title{ \vbox{ \hbox{MIFP-03-06\ \ \ USC-03/01} 
\hbox{\tt hep-th/0304132} }} {\vbox{\vskip -1.0cm
\centerline{\hbox
{A Dielectric Flow Solution with Maximal Supersymmetry}}
\vskip 8 pt
\centerline{
\hbox{}}}}
\vskip -0.9cm
\centerline{C.N.\ Pope$^*$ and N.P. \ Warner$^\dagger$} 
\bigskip\bigskip
\centerline{$^*${\it George P. \& Cynthia W. Mitchell Institute for 
Fundamental Physics}} 
\centerline{{\it  Physics Department, Texas A\&M University}} 
\centerline{{\it College Station, TX
77843-4242, USA}}
\bigskip
\medskip
\centerline{$^\dagger$ {\it Department of Physics and Astronomy}} 
\centerline{{\it University of Southern California}} 
\centerline{{\it Los Angeles, CA
90089-0484, USA}} 

\vskip 1.0cm
\centerline{{\bf Abstract}}
\medskip
We obtain a solution to eleven-dimensional supergravity that
consists of $M2$-branes embedded in a dielectric distribution
of $M5$-branes.  Contrary to normal
expectations, this solution has maximal  supersymmetry  for a 
brane solution ({\it i.e.} sixteen supercharges).  While the
solution is constructed using gauged supergravity in four dimensions,
the complete eleven-dimensional solution is given.  In particular, we obtain
the Killing spinors explicitly, and we find that they are characterised
by a duality rotation of the standard Dirichlet projection matrix
for $M2$-branes.

\vskip .3in
\Date{\sl {March, 2003}}

\parskip=4pt plus 15pt minus 1pt
\baselineskip=15pt plus 2pt minus 1pt

\newsec{Introduction}

The classification of supersymmetric backgrounds with tensor gauge
field fluxes remains a poorly understood subject, which we believe
will ultimately yield some beautiful and rich structures. These sorts
of backgrounds are essential to the study of holographic RG flows, and
yet we still do not know enough about them to construct some of the
most basic of holographic flows.  There are many examples, and some
simple, unifying ideas, like the ``harmonic rule,'' and more recently,
the application of $G$-structures, but the former has rather limited
application, and although the latter idea has led to some very
interesting insights, it is still under development (see, for example,
\refs{\GauntlettSC,
\GauntlettNW,\GauntlettFZ}).

There are some surprising  gaps in our knowledge.  For example,
there is the construction of the most general $\cN=2^*$ holographic flow 
from the  $\cN=4$ fixed point theory.  The Wilsonian effective action of
the field theory is known \DonagiCF, but the corresponding holographic flow is 
only known for one point on the (infinite dimensional) Coulomb branch \PilchUE.
There have been some recent examples that go beyond the usual 
harmonic rule, and go against the conventional wisdom in that they are 
combinations of branes and anti-branes, and yet preserve some
supersymmety (see, for example, \refs{\MateosQS,\BakXX}).   There have also been 
some recent  conjectures about  continuous duality symmetries that trade pure metric 
deformations  for $RR$-fluxes, while preserving supersymmetry \CorradoWX.

Our purpose here is to provide, and study in detail, an example that
touches upon all the foregoing issues.  Our example is unusual in that it
is a flow solution that has {\it sixteen} supersymmetries, and is not
merely some set of parallel ``harmonic'' branes.  Indeed, it looks
more like $M2$-branes embedded in a  dielectric distribution  $M5$-branes.  
Unlike previous examples, the solution presented here
preserves half the supersymmetries, and indeed it has the maximal
supersymmetry possible for a brane solution.  Thus our solution
provides an interesting extension of the conventional wisdom, and this
will need to be properly incorporated into any classification.
Indeed, since we will exhibit all the supersymmetries explicitly in
eleven dimensions, it would be interesting to see if the ideas of
$G$-structures could be used to understand this example more deeply,
and perhaps to generalise it.

Our solution is a little reminiscent of the solutions of  \BakXX, in
which supersymmetric combinations of $D2$ branes and anti-$D2$ branes
are generated by introducing electric and magnetic fields on the
world volume.  The result is a form of ``dielectric effect'' in which
$D0$ branes and fundamental strings are dissolved into the $D2$-brane,
making a composite $1/4$ supersymmetric system.  Here we present a
continuous family of solutions that starts with the standard,
near-brane, large $N$ limit of $M2$-branes, and then turns on a $3$-form
potential that appears to be sourced by a dielectric family of
$M5$-branes that contain the original $M2$-branes.
Turning on the additional $3$-form potential preserves the amount of
supersymmmetry, but the actual supersymmetry parameters rotate as the
parameter is changed.

Another motivation in studying this solution was a conjecture arising
out of the holographic description of flows from $\cN=2$ quiver gauge
theories.  In \CorradoWX\ it was argued that a particular subsector of
such gauge theories was holographically dual to a particular family of
gauged $\cN=4$ supergravity theories in five dimensions.  One
consequence of this work was the prediction of a global $SU(p+1,1)$ symmetry
that acts on the large $N$, $A_{p}$ quiver gauge theories.  In
particular, the $SU(p)$ subgroup of this symmetry acts on the
non-trivial $\cN=1$ superconformal fixed points, making them into a
${\Bbb C} {\Bbb P}^{p}$ fixed surface.  From the field theory point of
view such a symmetry is plausible (at large $N$), and from
the five-dimensional supergravity perspective, this symmetry is obvious \CorradoWX.
On the other hand, from the perspective of the underlying,
ten-dimensional IIB theory, such a symmetry involves trading
K\"ahler moduli of blown-up ${\Bbb P}^1$'s for topologically trivial
fluxes, while preserving supersymmetry.  Unfortunately, the
ten-dimensional solution seems to be too complicated for explicit
construction.  On the other hand, the general issue of ``rotating
metric deformations into fluxes'' is important, and we would like to
understand it more deeply; hence this example.

As with a number of interesting, and non-trivial examples, the origin
of this one is gauged supergravity.  In section 2 we use gauged
supergravity in four dimensions to obtain a one-parameter family of
flows that preserve sixteen supersymmetries.  These flows preserve an
$SO(4) \times SO(4)$ symmetry, and involve a four-dimensional scalar
and pseudo-scalar.  The parameter, $\zeta$, remains fixed along the
flow, and changing $\zeta$ rotates the scalar into the pseudo-scalar.
From the eleven-dimensional perspective of $M$-theory, a
four-dimensional scalar comes from the internal metric, while the
pseudo-scalar comes from the internal $3$-form potential,
$A^{(3)}_{mnp}$.  Thus a ``trivial'' rotation of the four-dimensional
configuration has a highly non-trivial effect in eleven dimensions.

  Fortunately, the uplift formulae are known
\refs{\deWitNZ,\deWitIY,\CveticAU}, and they have a particularly
simple form \CveticAU\ for the fields we consider here.\foot{
Strictly speaking, the uplift formulae for $\cN=8$ gauged supergravity 
obtained in \refs{\deWitNZ,\deWitIY} were not explicit in the
$4$-form sector, while the uplift formulae for $\cN=4$ gauged supergravity
obtained in \CveticAU \ were complete and explicit in the entire bosonic
sector, but did not include the fermions.  Thus our present work 
provides a non-trivial check on the uplift formulae, since we explicitly
verify that the flow solution in the four-dimensional gauged $\cN=4$ supergravity
does indeed exhibit the expected supersymmetry after uplifting to 
eleven dimensions.} In section 3, we use these results to obtain the corresponding
eleven-dimensional backgrounds.  In section 4 we compute explicitly
the sixteen supersymmetries in eleven dimensions.  In section 5 we
show how the $\zeta=0$ flow has a standard ``harmonic'' form, and
represents a distribution of $M2$-branes that have been spread out,
with constant density, in a $4$-ball in the ${\Bbb R}^4
\subset {\Bbb R}^8$ transverse to the branes.  The $\zeta=\pi$ flow is
similarly simple: The branes a spread uniformly in a $4$-ball in the
orthogonal ${\Bbb R}^4 \subset {\Bbb R}^8$.  In the latter part
of section  5, we examine
the solution for the intermediate values of $\zeta$.  The internal
gauge fields are reminiscent of the harmonic rule for $M5$-branes, and
they spread into ${\Bbb R}^4 \times {\Bbb R}^4 \subset {\Bbb R}^8$,
with one distribution of $M5$-branes in the first ${\Bbb R}^4$ factor, 
and another distribution of $M5$-branes in the second ${\Bbb R}^4$.  
We also show how changing $\zeta$ rotates the supersymmetry parameters.

\newsec{The flow solutions in four-dimensional supergravity}

Following \refs{\CremmerUP,\deWitIG}, we define the action of the  
$E_{7(7)}$ by
\eqn\Evars{\eqalign{ \delta\, z_{IJ} ~=~ \Sigma_{IJKL}\,
z^{KL} \cr \delta\, z^{IJ} ~=~ \Sigma^{IJKL}\,
z_{KL} \,,}}
where indices are raised and lowered by complex 
conjugation, and where one has
$$
\Sigma_{IJKL} ~=~  \overline{(\Sigma^{IJKL})}
~=~ \coeff{1}{24}\, \varepsilon_{IJKLPQRS}\, 
\Sigma^{PQRS} \,.
$$
In this formulation, we are going to consider a very simple
 $SL(2,\IR) \subset  E_{7(7)}$ defined by
\eqn\GPgens{ \Sigma_{IJKL} ~=~ 24 \, \big(\, z_0 \, 
\delta^{1234}_{[IJKL]} ~+~
\bar z_0\,\delta^{5678}_{[IJKL]} \,\big)  \,,}
where $z_0$ is a complex parameter corresponding to
the non-compact generators of the  $SL(2,\IR)$.    

This subset of the scalars has a manifest $SO(4)
\times SO(4)$ invariance acting on the indices
$1, \dots,4$ and $5,\dots,8$ separately.  This invariance
is a subgroup of the full $SO(8)$ gauge symmetry, and so
it will be an invariance of the solutions that we obtain here.

It is important to note that the $U(1)$ subgroup of the
$SL(2,\IR)$ is {\it not} a subgroup of the $SO(8)$ gauge symmetry:
This $U(1)$ rotates scalars into pseudo-scalars  in the four-dimensional
theory, and it is the $U(1)$ in the $SU(8) \subset  E_{7(7)}$ defined by
\eqn\Uone{ U ~=~ {\rm diag} (+ i \, \zeta,+ i \, \zeta,+ i \, \zeta,+ i \, 
\zeta, - i \, \zeta, - i \, \zeta, - i \, \zeta, - i \, \zeta) \,.}
This $U(1)$ is thus not an {\it a priori} symmetry of the solution.
However, it is a duality symmetry of a truncated form of the $\cN=8$
theory.

To be more explicit, if we truncate the entire $\cN=8$ theory to
the singlets of the second of the $SO(4)$ factors in
the $SO(4) \times SO(4)$ symmetry, then this will truncate the
$\cN=8$ supergravity down to $\cN=4$ gauged $SO(4)$ supergravity 
(in which the gauged $SO(4)$ is the first of the $SO(4)$ factors).
In this truncated theory, the $U(1)$ acts as an
``electric-magnetic'' duality symmetry on the gauge fields
of the supergravity, and is thus a symmetry of the equations
of motion.

One can use this $U(1)$ action to parametrise $z_0$:
\eqn\polcoords{z_0 = \coeff{1}{2}\, \alpha \, e^{ i \,\zeta}  \,.}
and then the scalar kinetic term takes the form
\eqn\scalkin{ (\partial_\mu \, \alpha)^2 ~+~ \coeff{1}{4}\,  
\sinh^2 2\alpha\, (\partial_\mu \, \zeta)^2 \,.}

On this sector the scalar potential is extremely simple, namely
\eqn\sugrpot{\cP ~=~ -{1\over L^2} \, (2  ~+~ \cosh 2\alpha ) \,,}
where we have replaced the usual supergravity gauge coupling,
$g$, according to
\eqn\gLreln{ g~=~ {1 \over \sqrt{2} \, L} \,.}

To find the supersymmetries one computes the $SU(8)$ tensor,
$A_1^{ij}$, that appears in the gravitino transformation rule.  One
finds that it is real and proportional to the identity matrix:
\eqn\Aonetens{ A_1^{ij} ~=~  \cosh \alpha  \, \delta^{ij} \,.}
This means that if we have any supersymmetry then we have 
maximal supersymmetry.  It also implies that there is a
superpotential,
\eqn\Wpot{ \cW  ~=~  \cosh\alpha  \,,}
and indeed one has
\eqn\PWreln{\cP ~=~ {1\over L^2}\, \bigg|{\del \cW \over \del 
\alpha}  \bigg|^2   ~-~  {3\over L^2}\, |\cW |^2 \,.}

The consequence of all this is that if we use the usual flow metric,
\eqn\RGFmetric{
ds^2_{1,3} = e^{2 A(r)} \eta_{\mu\nu}\,  dx^\mu dx^\nu + dr^2 \,,}
where $\eta_{\mu\nu}$ is the flat, Poincar\'e invariant metric
of the $M2$-brane, then the following equations of motion yield
a {\it maximally} supersymmetric flow:
\eqn\floweqs{ {d \alpha \over d r} ~=~  - {1 \over L}\,
{\del \cW \over \del \alpha}  \,, \qquad {d\zeta\over dr} ~=~ 0    \,, \qquad
{d A \over d r} ~=~ {1 \over L} \, \cW  \,.} 
These have the solution 
\eqn\flowsol{
e^\alpha ~=~ \coth{r\over 2 L}\,,\qquad 
e^A ~=~ \sinh{r\over L} ~=~{1 \over \sinh \alpha } \,,\qquad \zeta ~=~ {\rm const}\,.}
The remarkable fact is that the potential, the superpotential
and the  flows are all independent of the choice of $\zeta$, and
in particular the flow is maximally supersymmetric for all
choices of $\zeta$.  The $U(1)$ rotation by $\zeta$ is  only a duality symmetry of  
$\cN=4$ supergravity, and thus might, {\it a priori}, be expected to preserve
the four supersymmetries of that theory.  However, we have shown a stronger
result:  This duality symmetry preserves all the supersymmetries of the 
$\cN=8$ theory.    Again, we stress that rotations in $\zeta$ are
generically {\it not} symmetries of the gauged supergravity in four dimensions:
This rotation takes scalars into pseudo-scalars.  In the linearised
 eleven-dimensional  theory, such a rotation   takes metric modes
into tensor gauge field modes.  This symmetry is thus rather
unexpected, and must have interesting geometric consequences.

\newsec{Consistent Truncations/ $\cN=4$ supergravity}

Since our flow lies entirely within the $\cN=4$, $SO(4)$ gauged supergravity
theory, we can use the remarkably simple ``uplift'' formulae
of \CveticAU\ to find the corresponding solution in eleven dimensions.
We therefore review the results for this particular  Kaluza-Klein
reduction Ansatz for obtaining $\cN=4$ $SO(4)$ gauged supergravity in
four dimensions from an $S^7$ reduction of eleven-dimensional supergravity.  
For our present purposes we can set the Yang-Mills fields to zero,
since these do not 
participate in the four-dimensional supergravity solution that we are
considering here.  After doing this, the results for the consistent 
embedding obtained in \CveticAU\ are as follows.  The reduction 
Ansatz for the eleven-dimensional metric is given by
\eqn\metans{
d\hat s_{11}^2 = \Omega^2\, ds_4^2 + 2 g^{-2}\,
\Omega^2\, d\theta^2 + \ft12
g^{-2}\, \Omega^2\, \Big[\fft{c^2}{Y}\, \sum_i 
(\sigma^i)^2 + \fft{s^2}{\wtd Y}\, \sum_i 
(\td \sigma_i)^2\Big]\,, }
where
\eqn\defsone{\eqalign{
\tX ~\equiv~ &  X^{-1}\, q\,,\qquad q^2 ~\equiv~ 1 + \chi^2\, X^4\,,\qquad
  c ~\equiv~  \cos\theta \,,\qquad s~\equiv~ \sin\theta \cr
\Omega ~\equiv~ & \Big[Y \, \wtd Y \Big]^{\fft16} 
\,,\qquad Y ~\equiv~ (c^2\, X^2 + s^2)\,, \qquad 
\wtd  Y ~\equiv~(s^2\, \tX^2 + c^2) \,. }}
The constant $g$ is the supergravity gauge coupling constant.
The three quantities $\sigma_i$ are left-invariant 1-forms on
$S^3=SU(2)$, and the three $\td\sigma_i$ are left-invariant 1-forms
on a second $S^3$.    They satisfy\foot{Note that we have changed 
orientation conventions relative to those used in \CveticAU.}
\eqn\leftinv{
d\sigma_i = \ft12 \ep_{ijk}\, \sigma_j\wedge \sigma_k\,,\qquad
d\td\sigma_i = \ft12 \ep_{ijk}\, \td\sigma_j\wedge \td\sigma_k\,.}
Explicitly, in terms of Euler angles, $\varphi_i$, one has
\eqn\oneforms{\eqalign{\sigma_1 ~\equiv~&  \cos \varphi_3\, d\varphi_1 ~+~ 
\sin\varphi_3\, \sin\varphi_1\, d \varphi_2 \,, \cr
\sigma_2 ~\equiv ~&  \sin\varphi_3\, d\varphi_1 ~-~ 
\cos\varphi_3\, \sin\varphi_1\, d \varphi_2 \,, \cr
\sigma_3 ~\equiv ~&  \cos\varphi_1\, d\varphi_2 ~+~   d \varphi_3  \,,}}
with a similar expression for the $\td\sigma_i$ in terms of
a second set of Euler angles, $\td\varphi_j$.

The remaining
bosonic fields of the $\cN=4$ supermultiplet are the dilaton
$\phi$ and the axion $\chi$.  The dilaton parameterises the quantity
$X$ appearing in \metans\ and \defsone, being related to it by
$$
X= e^{\fft12\phi}\,.
$$

With the Yang-Mills fields set to zero, the reduction Ansatz for
$\hat F_\4$ is given by
\eqn\fansone{ \hat F_\4 = -g\, \sqrt2\, U\, \ep_\4 -
\fft{4s\, c}{g\,\sqrt2}\, X^{-1}\, {*dX}\wedge
d\theta + \fft{\sqrt2 s\, c}{g}\, \chi\, X^4\, {*d\chi}\wedge d\theta +
\hat F_\4'\,,}
where
$$
U = 1+Y + \wtd Y= X^2\, c^2 + \tX^2\, s^2  + 2 \,,
$$
and $\hat F_\4' = d\hat A_\3'$, with
\eqn\athreep{
\hat A_\3' = f\, \ep_\3 + \td f\, \td\ep_3\,.}
Here $\ep_\3 = \ft16 \ep_{ijk} \, \sigma_i\wedge \sigma_j\wedge \sigma_k$ and
 $\td\ep_\3 = \ft16 \ep_{ijk} \, \td \sigma_i\wedge \td \sigma_j\wedge 
\td \sigma_k$.
The functions $f$ and $\td f$ are given by
\eqn\ftf{\eqalign{
f = & -\fft1{2\sqrt2}\, g^{-3}\, c^4 \chi\, X^2\, (c^2\, X^2+s^2)^{-1}\,,\cr
\td f  = & \fft1{2\sqrt2}\,
g^{-3}\, s^4\, \chi\, X^2\,  (s^2\, \tX^2 + c^2)^{-1}\,.}}
The field strength contribution $\hat F_\4'$ is therefore given by
\eqn\fanstwo{\eqalign{
\hat F_\4' = & \fft{\del f}{\del\chi}\, d\chi\wedge \ep_\3 + \fft{\del
f}{\del X}\, dX\wedge \ep_\3 + \fft{\del f}{\del \theta}\, d\theta \wedge
\ep_\3  \cr 
  + & \fft{\del \td f}{\del\chi}\, d\chi\wedge \td\ep_\3 + \fft{\del
\td f}{\del X}\, dX\wedge \td\ep_\3 + \fft{\del \td f}{\del \theta}\, d\theta
\wedge \td\ep_\3\,.}}

To make contact with gauged supergravity we need to reparametrise
the $SL(2,\IR)/U(1)$
coset space in terms of its Lie algebra
generators.  The group manifold of $SL(2,\IR)$ can be thought of
as the hyperboloid
$$
X_0^2 ~+~ X_1^2 ~-~ X_2^2 ~=~ 1
$$
in $\IR^3$.  In the reduction Ansatz above, we have used horospherical 
coordinates in which one has
$$
X_0 ~+~ X_1 ~=~ e^\phi\,, \quad X_0 ~-~ X_1 ~=~ e^{-\phi} ~+~ e^{\phi}\,
\chi^2 \,, \quad X_2 ~=~ e^{\phi}\, \chi \,.
$$
If one goes to a complex ($SU(1,1)$) basis, and writes
the  non-compact Lie-algebra generator of $SL(2,\IR)$ as
in \polcoords, then
one has $X_0 = \cosh 2\alpha,\ X_1 = \sinh 2\alpha\, \cos\zeta$
and $X_2 = \sinh 2\alpha\, \sin\zeta$, from which one obtains
\eqn\coordtrf{e^\phi ~=~  \cosh 2 \alpha + \sinh 2\alpha \, \cos\zeta\,,
\qquad \chi ~=~  {\sinh 2 \alpha\, \sin\zeta \over 
\cosh 2\alpha + \sinh 2\alpha\, \cos\zeta}\,.}

    With these changes of variable, and using \gLreln, the metric may be 
written in terms of the frames
\eqn\frames{\eqalign{  e^1 ~=~& \Omega\, e^A\, dt \,, \qquad
e^2 ~=~   \Omega\, e^A\, dx \,, \qquad e^3 ~=~  \Omega\, 
e^A\, dy \,, \cr \qquad  e^4 ~=~&  \Omega\,  dr \,, \qquad 
e^5 ~=~ 2\, L\,  \Omega\, d\theta  \,,  \cr
\qquad e^{j+5}  ~=~&   L \,  \Omega\, Y^{-{1 \over2}}\,
\cos \theta \, \sigma_j \,,  \qquad   e^{j+8}  ~=~ L \, \Omega\, 
\wtd Y^{-{1 \over2}}\, \sin \theta \, \td \sigma_j  \,, \qquad j=1,2,3 \,,}}
where the functions $Y$ and $\wtd Y$ are given by 
\eqn\fundefns{\eqalign{Y ~=~&  \sin^2 \theta +  \cos^2 \theta \,  (
\cosh 2\alpha + \cos\zeta \, \sinh 2 \alpha  )  \,, \cr
\wtd Y ~=~&  \cos^2 \theta +  \sin^2 \theta \,  (
\cosh 2\alpha -  \cos\zeta \, \sinh 2\alpha  )\,.}}

The metric defined by \frames\ has a manifest symmetry of $SO(4)
\times SO(4)$, and there is also an interchange symmetry
\eqn\discsymm{\theta \to {\pi \over 2} - \theta \,, \qquad
\alpha \to - \alpha \,.}

The $3$-form potential is given by
\eqn\Aansatz{\eqalign{A^{(3)} ~=~ &  
{k^3 \, Z \over \sinh^3 \alpha } \, dt
\wedge dx \wedge dy  \cr & ~+~  \, L^3\,  \sin\zeta  \,  
\sinh 2\alpha  \, 
\bigg( {  \cos^4 \theta  \over Y}  \  \sigma_1 \wedge \sigma_2 \wedge 
\sigma_3  ~-~ 
{  \sin^4 \theta  \over \wtd Y}  \  \td \sigma_1 \wedge \td \sigma_2 
\wedge \td \sigma_3  \,
\bigg) \,,}}
where the function $Z$ is defined by
\eqn\Xthreedefn{ Z ~=~   {1 \over 2\, \cosh \alpha} \, (Y + \wtd Y)  ~=~ 
\cosh \alpha +  \cos\zeta  \, \sinh \alpha \,\cos 2\theta \,. }

This configuration satisfies  equations of motion
\eqn\eqnmot{ R_{MN} ~+~ R \, g_{MN} ~=~\coeff{1}{12}\, 
F^{(4)}_{MPQR}\, F^{(4)}_N{}^{PQR}\,,\qquad 
d*F^{(4)} ~=~ \coeff12 \, F^{(4)}\wedge F^{(4)}\,, }
where $*$ is defined using $\epsilon^{1\cdots 11} = 1$.

\newsec{Supersymmetry}

The gravitino variation is
\eqn\gravvar{\delta \psi_\mu ~=~  \nabla_\mu \, \varepsilon ~+~
\coeff{1}{288} \, 
\big( \,{\Gamma_\mu}^{\nu\rho\lambda\sigma}  ~-~ 8\, \delta^\nu_\mu \,
\Gamma^{\rho \lambda \sigma} \,  \big )  \, F_{\nu \rho \lambda \sigma} \, 
\varepsilon \,.}
and we will take the gamma-matrices to be
\eqn\gammamats{\eqalign{& \Gamma_1 ~=~  -i \, \Sigma_2 \otimes 
\gamma_9 \,, \quad
\Gamma_2~=~  \Sigma_1 \otimes \gamma_9 \,, \quad 
\Gamma_3~=~   \Sigma_3 \otimes \gamma_9 \,, \cr &
\Gamma_{j+3} ~=~  {\cal I}_{2 \times 2} \otimes \gamma_j \,, 
\quad j=1,\dots,8 \,,}}
where the $\Sigma_a$ are the Pauli spin matrices, $\oneone$ is
the Identity matrix, and  the $\gamma_j$ are real, symmetric $SO(8)$
gamma matrices.  As a result, the $\Gamma_j$ are all real, with 
$\Gamma_1$ skew-symmetric and $\Gamma_j$ symmetric for $j>2$.  One also
has:
$$
\Gamma^{1 \cdots\cdots 11} ~=~ \oneone \,,
$$
where $\oneone$ will henceforth denote the $32 \times 32$ identity matrix.

    Poincar\'e invariance parallel to the brane means that we can choose
a frame in which the Killing spinors are independent of $(t,x,y)$.  It
is instructive to start with the supersymmetry variations $\Gamma^1\, 
\delta\psi_1 = \Gamma^2\,  \delta\psi_2=\Gamma^3\,  \delta\psi_3=0$.
These equations are identical, and reduce to a single condition of the
form
\eqn\genform{
\Big(x_1\, \Gamma^4 + y_1\, \Gamma^5 - y_2\, \Gamma^{1234} + 
x_2\, \Gamma^{1235} + x_3\, \Gamma^{4678} + y_3\, \Gamma^{5678}
+ x_4\, \Gamma^{49\, 10\, 11} + y_4\, \Gamma^{59\, 10\, 11}\Big)
\varepsilon ~=~0\,.}
One can easily check that there are 16 solutions, if and only if 
\eqn\conifold{
\vec x\cdot \vec x = \vec y\cdot \vec y\,,\qquad \vec x\cdot \vec y =0\,,}
and otherwise there are no solutions.  It is amusing to note
that \conifold\ defines the conifold.  Remarkably, it turns out that the
solutions to \genform\ all yield solutions to the complete set of
Killing-spinor conditions. The matrix in \genform\ thus 
determines the entire family of solutions.

   In practice, it is considerably simpler, and indeed more enlightening,
to reduce the problem further by first defining
\eqn\Mops{ \cM_\mu \, \varepsilon ~=~ \Gamma^\mu \,(\delta \psi_\mu  ~-~
\partial_\mu  \varepsilon) \,,
\quad {\it with\ no\ sum\ on} \  \mu \,,}
and then observing that the projection matrices
\eqn\projs{\eqalign{ {\Pi}_1^\pm ~=~& \coeff{1}{2}\, 
\big ( \oneone ~\pm~ \Gamma^{12}
\big)\,, \qquad  {\Pi}_2^\pm~=~\coeff{1}{2}\, 
\big ( \oneone ~\pm~ \Gamma^{679\,10}
\big)\,, \cr    {\Pi}_3^\pm~=~& \coeff{1}{2}\, 
\big ( \oneone~\pm~ \Gamma^{68 9 \,11}
\big) \,, \qquad {\Pi}_4^\pm~=~  \coeff{1}{2}\, 
\big (  \oneone ~\pm~ \Gamma^{78 \,10\,11}
\big)\,,}}
all commute with the explicit forms of the operators ${\cM_\mu}$.
The projector ${\Pi}_4$ is redundant given ${\Pi}_2$ and ${\Pi}_3$, 
and the effect  of these projectors is to reduce each of the 
gravitino variations
to  a set of eight  $4 \times 4$ matrices.  The Killing
spinors then form two-dimensional eigenspaces of each of these eight 
matrices.  
 
    The solutions are most easily characterised in the following manner.
Introduce the following functions:
\eqn\fndefns{\eqalign{h_0 ~\equiv~& {1\over 2 \, \cosh \alpha} \,  
\bigg( \sqrt{Y \over \wtd Y} ~+~  \sqrt{\wtd Y \over Y} \bigg) \,, \cr 
h_1~\equiv~&\sin\zeta \, \sinh\alpha\,   {\cos\theta 
\over \sqrt{Y}} \,, \qquad   h_2~\equiv~ \sin\zeta \, \sinh\alpha\,   
{\sin\theta \over \sqrt{\wtd Y}} \,, }}
and define the matrix
\eqn\PMat{ {\cal P}~\equiv~ \coeff{1}{2}\, \Big(\oneone~+~ 
h_0 \, \Gamma^{123}     ~+~ \Gamma^{45}\, \big(h_2 \, \Gamma^{678}~+~ 
 h_1 \,\Gamma^{9\,10\,11} \,\big) \Big)\, \,. } 
The functions, $h_j$, satisfy the identity
\eqn\hident{h_0^2 ~+~ h_1^2 ~+~ h_2^2 ~=~ 1 \,,}
and hence $\cP$ is a projection matrix, satisfying ${\cal P}^2 ~=~{\cal P}$.  
The solutions  to $\delta \psi_\mu =0$ for $\mu =1,2,3$ are then
given by the solutions to
\eqn\condone{ {\cal P} \, \varepsilon_0  ~=~0 \,.}

The gravitino variations parallel to the sphere are largely controlled
by the symmetries.  The only subtle issue is the $SU(2)$ helicity
projectors, and for this it is useful to recall that the situation for
constant spinors on $\IR^4$, but written in terms of polar
coordinates.  That is, consider the covariantly constant spinors
\eqn\Rfourspin{ \nabla_\mu \, \varepsilon ~=~ 0 \,,}
for the metric
\eqn\Rfourmet{ds^2  ~=~ dr^2 ~+~ \coeff{1}{4} \, r^2\, 
\big(\sigma_1^2+\sigma_2^2 +
\sigma_3^2 \big)  \,,}
One finds that \Rfourspin\ may be written
\eqn\Rfourspin{dx^\mu \,  \nabla_\mu \, \varepsilon ~=~  d\, \varepsilon ~-~ 
\coeff{1}{8} \,
\sigma_k \, \eta^{k}_{ab}\, \gamma^{ab}\, \varepsilon ~=~ 0 \,,}
where the $\eta^{k}_{ab}$ are the self-dual 't Hooft matrices.  The
connection term in \Rfourspin\ thus acts on only one spinor helicity.
Thus the solution to \Rfourspin\ are spinors that are of one helicity
and are independent of the coordinates of $S^3$, and spinors of the
other helicity that are the non-trivial solutions of \Rfourspin, but
whose dependence on angle can be obtained by using Lie transport of
the spinor.

For the more complicated problem at hand we need to identify the
correct helicity projector, and the rest is straightforward.  To this
end, define matrices ${\cal E}$ and $\widehat {\cal E}$ by
\eqn\EMats{ \eqalign{ 
{\cal E}~\equiv~ & (h_1^2 + h_2^2)^{-{1 \over 2}}\,(\cosh^2\alpha \, 
\sin^2 2\theta + 
\cos^2 2\theta )^{-{1 \over 2}} \cr &\qquad \big( h_2\, \Gamma^{678}~+~ 
h_1 \,\Gamma^{9\,10\,11} \big)~ \big( \cosh \alpha \, \sin  2\theta\, 
\Gamma^{4}~+~
\cos  2\theta  \,\Gamma^{5} \big)\,, \cr
\widehat {\cal E}~\equiv~ & { \sin\theta \over \sqrt{Y}}\, \Gamma^{5678} ~-~ 
{ \cos\theta \over \sqrt{\wtd Y}}\, \Gamma^{5 9\,10\,11}   ~-~ 
{ \sin\zeta\, \sinh 2\alpha \, 
\sin\theta\, \cos\theta \over \sqrt{Y\, \wtd Y}}\, \Gamma^{5}\,.}}
These matrices satisfy
\eqn\matidents{  {\cal E}^2 ~=~  \widehat{\cal E}^2 ~=~\oneone\,, 
\qquad  \big[ {\cal E} \,,  {\cal P} \big]  ~=~ 0 \,,  
\qquad \big[ \widehat{\cal E} \,,  
{\cal P} \big]  ~=~ 0\,, \qquad {\cal E} \, \varepsilon_0 ~=~  
\widehat{\cal E} \, \varepsilon_0 \,,}
where $\varepsilon_0$ is a solution to \condone.  That is, both
square to the identity, commute with the projector, ${\cal P}$ of  \PMat,
and they have  equivalent   actions of spinors that 
satisfy \condone.  We will only need one of the matrices, but we include
the equivalent forms for completeness.

Introduce the helicity projectors, and helicity components of a spinor
\eqn\helproj{ \Pi_0^\pm ~\equiv~   \coeff{1}{2}\, 
\big (\oneone \pm {\cal E}\big) \,,
\qquad \varepsilon^\pm ~\equiv~  \Pi_0^\pm\, \varepsilon }

One can then check that the gravitino variations parallel to the spheres
reduce to the condition
\eqn\spherevars{\eqalign{ d\, \varepsilon ~+~  \coeff{1}{2}\, 
\Big( \big(   \sigma_1\,
 \Gamma^{78} ~-~  &\sigma_2\, \Gamma^{68}~+~ \sigma_3\,  \Gamma^{67}\big) \, 
\cr  ~+~  \big( & \td \sigma_1\, \Gamma^{10\,11} ~-~  \td \sigma_2\, 
\Gamma^{9\,11}~+~ 
\td \sigma_3\, \Gamma^{9\,10}\big) \Big)\,\Pi_0^-
 \, \varepsilon ~ ~=~ 0 \,.}}
In arriving that this equation we have also made use of \condone:  We assume,
henceforth that all our spinors indeed satisfy \condone.

Since $ \Pi_0^-\, \varepsilon^+ =0 $, it follows that such spinors solve the
variation equations parallel to the spheres if and only if
\eqn\rotsolplus{ \partial_\mu \, \varepsilon^+  ~=~  0\,, 
\qquad \mu =6,\dots,11\,.}
That is, the  $\varepsilon^+$ must be independent of  the coordinates on
 the spheres.  

The spinors, $\varepsilon^-$, of the opposite helicity also provide simple 
solutions. Define
\eqn\rotmat { \cR_{AB}(\varphi) ~=~ \cos(\coeff{1}{2} \, \varphi ) \,
\oneone_{AB} ~-~ \sin(\coeff{1}{2} \, \varphi ) \, \Gamma^{AB} \,,} 
and let
\eqn\grot{ g~\equiv~  \cR_{67}(\varphi_3)\, \cR_{68}(\varphi_1) \, 
\cR_{67}(\varphi_2) \,
 \cR_{9\,10}(\td\varphi_3)\, \cR_{9\,11}(\td\varphi_1) \, 
\cR_{9\,10}(\td\varphi_2) \,.}
By construction one has
\eqn\rtinv{ dg \, g^{-1} ~=~ - \coeff{1}{2} \Big( \big(   \sigma_1\,
 \Gamma^{78} ~-~   \sigma_2\, \Gamma^{68}~+~ \sigma_3\,  
\Gamma^{67}\big) \, ~+~
 \big(  \td \sigma_1\, \Gamma^{10\,11} ~-~  \td \sigma_2\, \Gamma^{9\,11}~+~ 
\td \sigma_3\, \Gamma^{9\,10}\big) \Big)  \,, }
and therefore the solution to the gravitino variation parallel to the
spheres  is generated by
\eqn\rotsolminus{\varepsilon^- ~=~  g\,  \Pi_0^-\, \varepsilon_0 \,, }
for some spinor, $\varepsilon_0$ that satisfies \condone,  and depends
only upon $r$ and $\theta$.

The last step is to determine the $r$ and $\theta$ dependence, and
solve the corresponding gravitino variation equations. 
This turns out to be elementary:  It is a theorem 
that if the $\varepsilon_k$ are solutions to $\delta \psi_\mu =0$
for $\delta \psi_\mu$ given by  \gravvar, then the bilinears
\eqn\spbilins{K_{ij}^\mu ~\equiv~ \bar \varepsilon_i \, \Gamma^\mu \,
\varepsilon_j \,,}
must be Killing vectors.  This completely fixes the $r$ and $\theta$
dependence of the spinors defined above.  Indeed, it suffices to check
that the first three components, $K_{ij}^\mu$, $\mu=1,2,3$ are all
constants (translations parallel to the brane).  Having thus
normalised $\varepsilon^\pm$, one can then check that indeed all the
gravitino variations vanish for the sixteen spinors given by solving
\condone, and using
\helproj, \rotsolplus\ and \rotsolminus.

It is, perhaps, useful to include one or two more explicit details.
Recall that one can reduce the problem to $4 \times 4$ blocks using
the projectors,
\projs.  In a suitably chosen basis,  these $4 \times 4$ blocks each
yields two orthonormal solutions of the form \condone:
\eqn\spsols{\eqalign{ \varepsilon_1~=~ & {1 \over \sqrt{2(h_0 +1)}} \, 
\left( \matrix{ h_0+1, & 0 ,& h_1, & h_2 \cr}\right) \,, \cr
\qquad  \varepsilon_2~=~ & {1 \over \sqrt{2(h_0 +1)}} \, 
\left( \matrix{0, &h_0+1,  
& h_2, & -h_1  \cr}\right) \,.}}
They are orthonormal in that $\varepsilon_1\cdot \varepsilon_1= 
\varepsilon_2\cdot \varepsilon_2 =1$, and $\varepsilon_1\cdot 
 \varepsilon_2 =0$. (One uses \hident\ to verify this.)   One also
has $\varepsilon_2 = \Gamma^{45}  \varepsilon_1$ for this pair of spinors.  

One can then apply $\Pi^\pm_0$ to these, and this generates
$\varepsilon^\pm$ as independent linear combinations of
$\varepsilon_1$ and $\varepsilon_2$.  One must once again normalise
these projected spinors.  The result is algebraically messy, but it
yields the Killing spinors.  One can then get the other fourteen
supersymmetries either by doing this in each of the $4 \times 4$
blocks, or simply by acting with $\Gamma^{23},
\Gamma^{67},\Gamma^{78}$ and $\Gamma^{68}$ on the $\varepsilon^\pm$
obtained in the first $4 \times 4$ block.

There is also a rather simple way to obtain the matrix $\cal E$:  There
is a linear combination of gravitino variations from which the tensor 
gauge field, $F$, cancels completely, leaving only connection terms.
Specifically, one has
\eqn\magic{\eqalign{\cM_1 + \cM_6 + \cM_9 ~=~ \coeff{1}{4L} \, \Omega^2 \, 
\Big(
\big( &h_2\,  \Gamma^{678}~+~ h_1 \,\Gamma^{9\,10\,11} \big)\cr & ~+~  
\big( \cosh \alpha  \, \sin  2\theta\, \Gamma^{4}~+~ \cos  2\theta  \,
\Gamma^{5}
 \big)\Big) \, \varepsilon \,,}}
for the matrices defined in \Mops.  This must vanish for the spinors
that are independent of the coordinates on the spheres, and indeed the
left-hand side of \magic\ is a multiple of $ \Pi_0^- \equiv
\coeff{1}{2}\, (\oneone - {\cal E} )$.  From this one can obtain
$\cal E$.

\newsec{Dielectric Rotations of the Coulomb Branch Flow}

\subsec{The Coulomb branch}

For $\zeta =0$, the flow defined in the previous sections lies
in the purely scalar sector of supergravity, and must reduce to a standard
``Coulomb branch flow,''  for which everything can be expressed in terms
of a single harmonic function.

Setting $\zeta =0$ simplifies things  significantly: The transverse
components of $A^{(3)}_{mnp}$ vanish, and one has 
$\wtd Y$ = $e^{-2 \,\alpha} Y$.
The pre-factor in front of the $M2$-brane section of the metric reduces to
$H^{-2/3}$, where
\eqn\harmfn{ H~\equiv~ { e^{ \alpha} \, \sinh^3 \alpha   \over Y |_{\zeta=0}}
~=~  { \sinh^3 \alpha   \over ( e^{ -\alpha} \,   \sin^2 \theta ~+~ 
 e^{ \alpha} \,   \cos^2 \theta) }\,. }
This is the harmonic function, but the metric is in some peculiar coodinates
adapted to gauged supergravity.  

  We therefore introduce new variables, according to 
\eqn\newvars{  x~=~ {1 \over \sqrt{ (e^{2 \,\alpha} - 1)}} \, \cos \theta \,,
\quad y~=~ {e^ \alpha  \over \sqrt{ (e^{2 \,\alpha} - 1)}} \, \sin \theta  \,.}
In terms of these, the metric becomes
\eqn\harmmet{\eqalign{ds_{11}^2 ~=~ H^{-2/3} k^2  \, (  -dx_0^2 + & dx_1^2 + 
dx_2^2)  \cr    + ~& 8\, H^{1/3} \, L^2\, 
\bigg( dx^2 +  dy^2 + \coeff{1}{4}\, x^2 \,
\sum_{j=1}^3 \, \sigma_j^2 + \coeff{1}{4}\, y^2 \, \sum_{j=1}^3 \, 
\td \sigma_j^2 \bigg)\,.}}
This is, of course, the requisite ``harmonic'' form, with flat metrics 
inside the parentheses.

In the new variables, the harmonic function takes the rather unedifying form
\eqn\newH{ H~=~  {1\over 16 \, x^2 \, y^2 \, (x^2 + y^2) \, \sqrt{\Delta}}  
 \,  \Big( (x^2 + y^2)\,\sqrt{\Delta} ~-~ \big( (x^2 + y^2)^2   + 
(x^2 -  y^2) \big)  \Big)  \,. } 
where
\eqn\Deltadefn{\Delta ~\equiv~  (x^2 + (y-1)^2)\, (x^2 + (y-1)^2) \,.}
However, by looking at the asymptotics of this for small $x$, one finds that
$$
H ~\sim~ {1 \over 4\, x^2} \,, \qquad {\rm for} \ \ y^2 < 1 \,.
$$
This means that the $M2$-branes are spread out into a solid $4$-ball,
defined by $y^2 <1$,  with a constant density of branes throughout the ball.
In particular, the function, $H(x,y)$ is given by
\eqn\Gfnform{H(x,y)~=~  {\rm const. }  \, \int_{z^2 <1} \, { d^4z \over 
(x^2  ~+~ (\vec y - \vec z)^2)^3 } \,,}
where $\vec y$ and $\vec z$ are vectors in $\IR^4$.

The Killing spinor analysis simplifies in that one now has
\eqn\newPMat{ {\cal P}~\equiv~ \coeff{1}{2}\, \big(\oneone~+~ 
 \Gamma^{123}  \big)\, \,. } 
If one writes $x = u\, \cos\psi$, $y = u\, \sin \psi$, then the matrix
$\widehat {\cal E}$ becomes, in these new coordinates, 
\eqn\newEMat{\widehat {\cal E}~\equiv~  \sin\psi \ \Gamma^{5678} ~-~ 
\cos\psi \ \Gamma^{5 9\,10\,11}    \,.} 
It is not completely trivial because the flat metric on $\IR^8$ metric has been
written in terms of angular  coordinates.  The earlier comments about
normalising the Killing spinors amount to putting the proper power of
the harmonic function into the spinor.

\subsec{Dielectric rotations}

The general solution presented in sections 2 and 3 has a very non-trivial 
form, and yet it is related via a simple $U(1) \subset SU(8)$ duality
rotation,  \Uone, to the much simpler, and well-understood ``harmonic''
M2-brane distribution described above.  Thus, in terms
of the eleven-dimensional background, the $SU(8)$ duality
rotation,  \Uone,  has a highly non-trivial effect that
we will now attempt to characterise.

Under the $U(1)$ rotation, the harmonic function \harmfn\ appears to
split into two components, which we will denote by $H_1$ and $H_2$,
with $H_j \sim Y_j^{-1}$.  These component functions do not seem to be
harmonic within any obvious flat metric, and so we have not defined
them precisely, but one would like to choose them so that
\eqn\harminterp{ {\Omega^2 \over \sinh^2(\alpha)} ~=~ (H_1\, H_2)^{-1/3}\,.}
Note that the left-hand side of \harminterp\ is the coefficient of the
$M2$-brane sections of the metric defined by \frames.  Also note that
this is consistent with \harmfn\ for $\zeta =0$.  These ``pseudo-harmonic''
functions also appear in the internal parts of the metric parallel to
the spheres.  Indeed, the eleven-metric has the schematic form
\eqn\metform{\eqalign{ds_{11}^2 ~\sim~ & (H_1\, H_2)^{-1/3} \, 
(  -dx_0^2 +  dx_1^2 + 
dx_2^2)  ~+~  h(r,\theta)  \big( dr^2 +  4\, L^2 \, d\theta^2) \cr  
 & ~+~ H_1^{2/3} 
\, H_2^{-1/3}  \, \Big(  \sum_{j=1}^3 \, \sigma_j^2 \Big) ~+~ 
H_1^{-1/3}  \, H_2^{2/3} \,
\Big( \sum_{j=1}^3 \, \td \sigma_j^2 \Big)\,.}} 
The way that the pseudo-harmonic functions appear is suggestive of
$M5$-branes wrapping the $3$-spheres, and intersecting over what was
the original $M2$-brane.  This suggestion is strongly supported by the
form of $A^{(3)}_{mnp}$: There is still an ``electric'' $M2$-brane
component parallel in the $123$ directions, while the components
parallel to the $678$ and $9\,10\,11$ directions are consistent with
$M5$-branes in the $1239\,10\,11$ and $123678$ directions
respectively. Note also that the pseudo-harmonic functions, and
{\it not} their inverses, appear as the components of $A^{(3)}_{mnp}$
parallel to the spheres.  We thus have a distribution of $M5$-branes
wrapping the spheres in the $679$ and $9\,10\,11$ directions.  The
relative sign for the corresponding components of $A^{(3)}_{mnp}$ in
\Aansatz\ make it tempting to think of these branes as being
$M5$-branes and anti-$M5$-branes, but such a distinction is largely a
matter of orientation.

This picture is particularly evident in the form of the 
Killing spinors.  The crucial step in finding the Killing spinors was to 
identify the projection matrix, $\cP$, that defines the relevant
sixteen-dimensional subspace in which they live.  It turns out that
the general projection, $\cP$, can be obtained via an $SU(8)$ rotation
of the simple projector, \newPMat, for the ``harmonic'' solution.

Define
\eqn\Drot{\cR ~=~  {1 \over \sqrt{2(h_0 +1)}} \, \Big( (h_0 +1) \, 
\oneone ~-~  h_1 \, \Gamma^{678} ~+~  h_2 \, \Gamma^{9\,10\,11} \Big) \,.}
One can then check that the projection matrix, $\cP$, of \PMat\ is related to 
the projection, $\half(\oneone +\Gamma^{123}) $,  of  \newPMat\ via
\eqn\rotharm{\cP ~=~ \coeff{1}{2}\,
\cR\, \big (\oneone ~+~  \Gamma^{123} \big)\, \cR^{-1} \,.}

The matrix, $\cR$, is a real rotation matrix in $SO(32)$.  Indeed,
since it only acts on the eight-dimensional ``internal space,'' we may
think of it as an $SO(16)$ matrix acting on the $8_s$ and $8_c$
spinors of the internal space.  Since $\cR$ involves the triple
products of gamma matrices, and these flip $SO(8)$ helicity, $\cR$
actually lives within the $SU(8)$ subgroup of $SO(16)$.  Since $\cR$
is an $SU(8)$ matrix that preserves $SO(4) \times SO(4)$ symmetry, and
rotates the Killing spinor projector from the harmonic flow to the
general flow, it is natural to think of $\cR $ as the
eleven-dimensional analogue of the $SU(8)$ rotation \Uone.  This
is precisely the local $SU(8)$ action that plays a crucial
role in the consistent truncation proof of \deWitIY.

The other non-trivial matrix needed to obtain the Killing spinors is
the helicity projector, $\cal E$.  The Killing spinors lie in either
singlets or doublets of the $(SU(2))^4$ symmetry of the solution, and
the role of $\cal E$ is to sort the solutions to \condone\ into linear
combinations that are either singlets or doublets of the $SU(2)$'s.
As remarked earlier, the form of $\cal E$ can be found by taking the
linear combination of gravitino variations from which the tensor gauge
field cancels.  However, it is also highly constrained, but not quite
uniquely determined by the fact that it must commute with the
rotation, $\cR$, and with the projector, $\cP$.  It may also be
shifted by multiples of $\cP$.

In terms of brane distributions, the effect of $\cR$ is quite dramatic.
The unrotated projector, \newPMat, is the standard Dirichlet projector
imposed by the presence of the branes.  Additional branes usually reduce
supersymmetry by imposing more projection conditions, however the dielectric 
distribution discovered here simply deforms the standard projector in
the directions of the fluxes generated by the dielectric $M5$-branes.

\newsec{Final Comments}

While there are many solutions that can be generated using gauged
supergravity theories, we find the solution presented here especially
interesting. It has the maximal amount of supersymmetry
possible for a brane solution, and yet consists of a complicated 
combination of $M2$-branes and $M5$-branes.  As was also seen
in \CorradoWX, the solution presented here shows how a relatively
simple duality symmetry of gauged supergravity can translate into 
a very non-trivial symmetry of the complete eleven-dimensional theory,
while preserving all the supersymmetries.
More generally, it is still an open problem to classify supersymmetric
compactifications in the presence of fluxes, and from this paper it
is evident that even maximally supersymmetric flows have a rather richer
structure than one might have otherwise expected.

\bigskip
\bigskip
\centerline{\bf Acknowledgements}
The authors would like to thank the {\it Isaac Newton Institute} in Cambridge, UK for
its support during the early stages of this work.  NW would also like to
thank the George P. \& Cynthia W. Mitchell Institute for Fundamental Physics
for hospitality while this work was completed.
This work was supported in part by funds
provided by the DOE under grant numbers DE-FG03-95ER-40917 and
DE-FG03-84ER-40168.
\vfill
\eject
\listrefs
\vfill
\eject
\end